\documentclass[journal]{IEEEtran}
\usepackage{graphicx,amssymb,amsmath}
\usepackage{multicol}
\usepackage[noadjust]{cite}
\usepackage{setspace}
\usepackage{stfloats}
\usepackage{midfloat}
\usepackage[normal]{threeparttable}
\usepackage{cuted}
\usepackage{cases,subeqnarray}
\usepackage{bm,multirow,bigstrut}
\usepackage{textcomp}
\usepackage{latexsym,bm}
\usepackage{booktabs,changebar}
\usepackage{xcolor}
\usepackage{mathtools}
\usepackage[linesnumbered,boxed]{algorithm2e}

\newtheorem{prop}{Proposition}

\IEEEoverridecommandlockouts
\begin{document}

\title{On the Spectral Efficiency of Space-Constrained Massive MIMO with Linear Receivers
\thanks{The work of J. Zhang and L. Dai was supported in part by the International Science \& Technology Cooperation Program of China (Grant No. 2015DFG12760), and China Postdoctoral Science Foundation (No. 2014M560081). The work of C. Masouros was supported by the Royal Academy of Engineering, UK, the Engineering and Physical Sciences Research Council (EPSRC) project EP/M014150/1.}}

\author{\authorblockN{Jiayi Zhang\authorrefmark{1}\authorrefmark{2}, Linglong Dai\authorrefmark{2}, Michail~Matthaiou\authorrefmark{3}, Christos Masouros\authorrefmark{4}, and Shi Jin\authorrefmark{5}\\
\authorblockA{\authorrefmark{1}School of Electronics and Information Engineering, Beijing Jiaotong University, Beijing 100044, P. R. China}\\
\authorblockA{\authorrefmark{2}Tsinghua National Laboratory for Information Science and Technology (TNList)\\
Department of Electronic Engineering, Tsinghua University, Beijing 100084, P. R. China}\\
\authorblockA{\authorrefmark{3}School of Electronics, Electrical Engineering and Computer Science, Queen's University Belfast, Belfast, U.K.}\\
\authorblockA{\authorrefmark{4}Department of Electronic and Electrical Engineering, University College London, Torrington Place, London, U.K.}\\
\authorblockA{\authorrefmark{5}National Mobile Communications Research Laboratory, Southeast University, Nanjing 210096, P. R. China}\\
 Email: jiayizhang@bjtu.edu.cn}}

\maketitle

\begin{abstract}
In this paper, we investigate the spectral efficiency (SE) of massive multiple-input multiple-output (MIMO) systems with a large number of antennas at the base station (BS) accounting for physical space constraints. In contrast to the vast body of related literature, which considers fixed inter-element spacing, we elaborate on a practical topology in which an increase in the number of antennas in a fixed total space induces an inversely proportional decrease in the inter-antenna distance. For this scenario, we derive exact and approximate expressions, as well as simplified upper/lower bounds, for the SE of maximum-ratio combining (MRC), zero-forcing (ZF) and minimum mean-squared error receivers (MMSE) receivers. In particular, our analysis shows that the MRC receiver is non-optimal for space-constrained massive MIMO topologies. On the other hand, ZF and MMSE receivers can still deliver an increasing SE as the number of BS antennas grows large. Numerical results corroborate our analysis and show the effect of the number of antennas, the number of users, and the total antenna array space on the sum SE performance.
\end{abstract}


\IEEEpeerreviewmaketitle
\section{Introduction}
As a disruptive technology for the fifth generation (5G) communication systems, massive MIMO has recently attracted extensive research and academic interest \cite{larsson2014massive,andrews2014will,zhang2015achievable,zheng2015large}. In massive MIMO systems, several co-channel users (UEs) simultaneously communicate with a BS equipped with a massive number of antennas (a few hundreds or even larger). Due to the deployment of a large antenna array, the channel vectors between the different UEs and the BS become asymptotically orthogonal \cite{zheng2015large}. Under this condition, dubbed as \textit{favorable propagation}, massive MIMO systems can achieve large array and spatial multiplexing gains by using simple linear signal processing methods at both the transmitter and receiver \cite{ngo2013energy}.

A critical issue pertaining to practical massive MIMO systems is the dense deployment of a massive number of antennas in a limited physical space. In general, if the inter-element spacing is more than half a wavelength, the communication channels can be considered as uncorrelated. However, for practical space-constrained massive MIMO systems, it is more likely that the antenna elements will be placed far less that half a wavelength apart. Under these conditions, the channel vectors for different UEs will not be asymptotically orthogonal. Therefore, a space-constrained massive MIMO architecture will suffer from increased spatial correlation, whose impact needs to be rigorously quantified and analyzed.

Numerous works have investigated the effect of spatial correlation on the performance of conventional MIMO systems with a relatively small number of BS antennas. The authors of \cite{matthaiou2011novel} presented upper and lower bounds on the achievable sum SE of MIMO systems with ZF receivers, especially over correlated Rayleigh and Ricean fading channels. In \cite{mckay2010achievable}, expressions for the exact achievable sum SE of MIMO with MMSE receivers were derived for correlated Rayleigh fading channels. In the context of massive MIMO systems, the authors of \cite{masouros2013large} approximated the performance of two distinct linear precoding schemes considering the spatial correlation at the transmitter. Recently, \cite{masouros2015space} demonstrated that, when the physical space is limited, the classical assumption of favorable propagation in massive MIMO systems is violated. However, only maximum ratio-transmission (MRT) precoding was considered in \cite{masouros2015space}. A lower bound on the achievable SE of uplink data transmission with MRC receivers at the BS was derived in \cite{ngo2013multicell}. In addition to information-theoretical studies, the authors of \cite{Teeti2015impact} investigated the impact of constrained space on the performance of subspace-based channel estimation schemes. To the best of our knowledge, there are no theoretical results on the SE of space-constrained massive MIMO with linear receivers, namely MRC, ZF and MMSE.

Motivated by the aforementioned considerations, we present a generic analytical framework for statistically characterizing the achievable SE of space-constrained massive MIMO with linear receivers. Specifically, the paper makes the following specific contributions:
\begin{itemize}
\item Motivated by some recent advances in the area of Wishart random matrix theory, we first present approximate expressions for the achievable sum SE of a massive MIMO system with MRC receivers. We show that a space-constrained antenna deployment will cause a saturation of the achievable sum SE with an increasing number of antennas for MRC receivers.
\item For ZF receivers, new upper and lower bounds on the achievable SE are derived, with the latter being particularly tight. We show that for uniform linear arrays, the achievable SE increases with the number of BS antennas $M$. Moreover, a larger number of UEs $K$ increases the sum SE of ZF receivers when $M \gg K$.
\item Finally, we derive an exact closed-form expression for the achievable SE, for MMSE receivers at the BS. Similar to ZF receivers, the sum SE of MMSE receivers also increases by deploying more BS antennas in space-constrained massive MIMO systems.
\end{itemize}
\emph{Notation:} In the following, $\bf{x}$ is a vector, and $\bf{X}$ is a matrix. We use ${\text{tr}}(\bf{X})$, ${\bf{X}}^T$, and ${\bf{X}}^H$ to represent the trace, transpose, and conjugate transpose of $\bf{X}$, respectively, while ${\tt{E}}\{\cdot\}$ denotes the expectation operator. The matrix determinant and trace are given by $|\bf{X}|$ and ${\text{tr}}(\bf{X})$, while ${{\bf{X}}_i}$ is ${{\bf{X}}}$ with the $i$th column removed. Finally, $[{\bf{X}}]_{ij}$ and ${\bf{x}}_{i}$ denote the $(i, j)$th entry and the $i$th column of $\bf{X}$, respectively.

\section{System and Channel Model}\label{se:model}
We consider the uplink of a single-cell massive MIMO system, where the BS with $M$ antennas simultaneously serves $K$ single-antenna UEs. The received vector ${\bf{y}} \in \mathbb{C}^{M\times 1}$ at the BS is given by
\begin{align}\label{eq:received_signal}
{\bf{y}} = \sqrt{p_u}{\bf{G}}{\bf{x}}+{\bf{n}},
\end{align}
where $p_u$ is the average power of each UE, ${\bf{x}}\in \mathbb{C}^{K\times 1}$ denotes the zero-mean Gaussian transmit vector from all $K$ UEs with unit average power, and the elements of ${\bf{n}}$ represent the additive white Gaussian noise (AWGN) with zero-mean and unit variance. The channel matrix between the BS and UEs can be written as ${\bf{G}} = {\bf{A}}{\bf{H}}{\bf{D}}^{1/2}$, where ${\bf{H}} \in \mathbb{C}^{P \times K}$ is the propagation response matrix standing for small-scale fading, and ${\bf{D}}  \in \mathbb{C}^{K \times K}$ denotes a diagonal matrix whose $k$th diagonal element $\zeta_k$ models the large-scale fading (including geometric attenuation and shadow fading) of the $k$th UE. We assume that large-scale fading changes very slowly such that all $\zeta_k$ are constant. Moreover, ${\bf{A}} \in \mathbb{C}^{M \times P}$ is the transmit steering matrix, with $P$ denoting a large but finite number of incident directions in the propagation channel \cite{masouros2013large}. For the sake of analytical simplicity, we assume that all UEs are seen from the same set of directions with cardinality $P$. Considering the widely used uniform linear antenna array, we can write ${\bf{A}}$ as \cite{ngo2013multicell,zi2014multiuser}
\begin{align}\label{eq:steering_matrix}
{\bf{A}} = {[{\bf{a}}\left(\theta_1\right),{\bf{a}}\left(\theta_2\right),\dots, {\bf{a}}\left(\theta_P\right) ]},
\end{align}
where ${\bf{a}}(\theta_i)$, for $i=1,2,\dots,P$ denotes a length-$M$ normalized steering vector as
\begin{align}\label{eq:steering_vector}
{\bf{a}}\left(\theta_i\right)= \frac{1}{\sqrt{P}}\left[1, e^{-j\frac{2\pi d}{\lambda}\sin{\theta_i}},\dots,e^{-j\frac{2\pi d}{\lambda}{(M-1)}\sin{\theta_i}}\right]^{T},
\end{align}
where $d$ is the antenna spacing, $\lambda$ denotes the carrier wavelength, and $\theta_i$ represents the direction of arrival (DOA). The normalized total antenna array space $d_0$ at the BS can be expressed as $d_0=\frac{dM}{\lambda}$. In \eqref{eq:steering_vector}, we use the factor $\frac{1}{\sqrt{P}}$ to normalize the steering vector ${\bf{a}}\left(\theta_i\right)$.

A key property of massive MIMO systems is that simple linear signal processing become near-optimal, while keeping the implementation complexity at very low levels \cite{zheng2015large}. Thus, we will hereafter consider the performance of space-constrained massive MIMO systems with linear receivers. We further assume perfect CSI is available at the BS \cite{ngo2013energy}. The linear receiver matrix ${\bf{T}} \in \mathbb{C}^{M \times K}$ is used to separate the received signal into $K$ streams by
\begin{align}\label{eq:receivers}
{\bf{r}} = {\bf{T}}^H {\bf{y}}= \sqrt{p_u}{\bf{T}}^H{\bf{G}}{\bf{x}}+{\bf{T}}^H{\bf{n}}.
\end{align}
Then, the $k$th element of the received signal vector, which corresponds to the detected signal for $k$th UE, is given by
\begin{align}
{r}_k = \sqrt{p_u}{\bf{t}}_k^H{\bf{g}}_k{x_k} + \sqrt{p_u} {\sum_{l\neq k}^K {\bf{t}}_k^H{\bf{g}}_l{x_l}} + {\bf{t}}_k^H{\bf{n}}.
\end{align}
Assuming that channel fading is ergodic, the achievable uplink SE, $R_k$, of the $k$th UE is given by \cite{ngo2013energy}
\begin{align}\label{eq:achievable_SE}
R_k = {\tt{E}}\left\{\log_2\left( 1+ \frac{{p_u}|{\bf{t}}_k^H{\bf{g}}_k|^2}{ {p_u}{\sum_{l\neq k}^K |{\bf{t}}_k^H{\bf{g}}_l|^2+\|{\bf{t}}_k\|^2}} \right)\right\}.
\end{align}
The uplink sum SE can be then defined as
\begin{align}\label{eq:achievable_SE_sum}
R = \sum\limits_{k = 1}^K {{R_k}}\;\;\; \text{in bits/s/Hz}.
\end{align}
In the following three sections, we analyze the achievable sum SE of space-constrained massive MIMO systems with different linear receivers, namely MRC, ZF, and MMSE, respectively.

\section{MRC Receivers}\label{se:MRC}
For the case of MRC receivers, we have ${\bf{T}}={\bf{G}}$ \cite{zhang2016spectral}. From \eqref{eq:achievable_SE}, the uplink SE for the $k$th UE boils down to
\begin{align}\label{eq:achievable_SE_mrc}
{R_k^{\text{MRC}}} = {\tt{E}}\left\{\log_2\left( 1+ \frac{{p_u}\|{\bf{g}}_k\|^4}{ {p_u}{\sum_{l\neq k}^K |{\bf{g}}_k^H{\bf{g}}_l|^2+\|{\bf{g}}_k\|^2}} \right)\right\},
\end{align}
where
\begin{align}\label{eq:g_k}
{\bf{g}}_k={\sqrt{\zeta_k} }{\bf{A}}{{\bf{h}}_k}.
\end{align}
We now present an approximation on the achievable sum SE of MRC receivers in the following proposition.
\begin{prop}\label{prop:mrc}
For space-constrained massive MIMO systems with MRC receivers, the approximated sum achievable SE is given by
\begin{align}\label{eq:achievable_SE_appro_result}
{{ {R}} ^{\text{MRC}}} \approx \sum\limits_{k = 1}^K {\log _2}\left( {1 + \frac{{{p_u}\left( {{M^2} + \sum\limits_{i = 1}^P {\beta _i^2} } \right)}{\zeta_k}}{{{p_u}\sum\limits_{l\neq k}^K {\zeta_l} {\sum\limits_{i = 1}^P {\beta _i^2} }{}  + M{\zeta_k} }}} \right),
\end{align}
where $\beta _i$ is the $i$th eigenvalue of the matrix ${{{\bf{A}}^H}{\bf{A}}}$.
\end{prop}
\begin{IEEEproof}
See Appendix \ref{se:A}.
\end{IEEEproof}

\newcounter{mytempeqncnt3}
\begin{figure*}[!b]
\normalsize
\setcounter{mytempeqncnt3}{\value{equation}}
\hrulefill
\vspace*{-4pt}
\setcounter{equation}{11}
\begin{align}\label{eq:achievable_SE_zf_lower_result}
{R }^{\text{ZF}} \geq {R_{\text{L}}^{\text{ZF}}} = \sum\limits_{k = 1}^{K}  {\log _2}\left( 1 + {p_u} {\zeta_k}\exp \left( {\sum\limits_{n \neq k}^K {\zeta_n}}\left({\psi \left( K \right) + \frac{{\left| {{{\bf{Y}}_{P - K + 1}}} \right|}}{{\prod\nolimits_{i < j}^P {\left( {{\beta _j} - {\beta _i}} \right)} }}}\right) -\left({\psi \left( n \right) + \frac{{\sum\limits_{n = P - K + 2}^P {\left| {{{\bf{Y}}_n}} \right|} }}{{\prod\nolimits_{i < j}^P {\left( {{\beta _j} - {\beta _i}} \right)} }}} \right) \right) \right),
\end{align}
\vspace{-6mm}
\setcounter{equation}{\value{mytempeqncnt3}}
\end{figure*}


Next, we provide numerical results to verify the analytical approximation in \eqref{eq:achievable_SE_appro_result}. Let us assume that the users are distributed uniformly at random in a hexagonal cell with a radius of 1000 meters, while the smallest distance between the UE to the BS is $  r_\text{min}= 100$ meters. Moreover, the pathloss is modelled as $r_k^{-v}$ with $r_k$ denoting the distance between the $k$th UE to the BS and $v=3.8$ denoting the path loss exponent, respectively. A log-normal random variable $s_k$ with standard deviation $8 $ dB is used to model shadowing. Combining these factors, large-scale fading can be given by $\zeta_k = s_k (r_k/r_\text{min})^{-v}$. We further assume $\theta_i$ are uniformly distributed within the interval $[-\pi/2,\pi/2]$.

\begin{figure}[htbp]
\centering
\vspace{-0mm}
\includegraphics[scale=0.6]{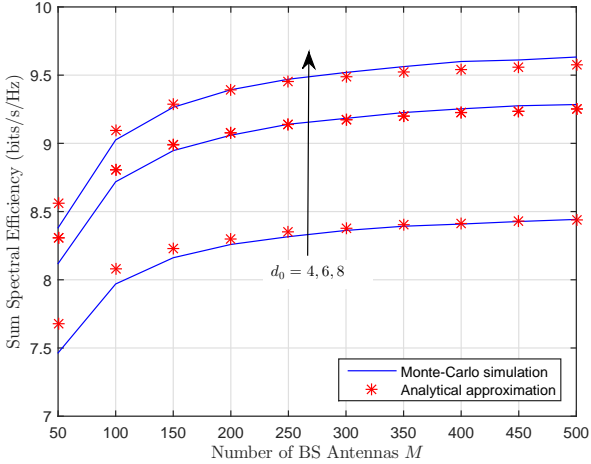}
\caption{Simulated and analytical approximation of the sum SE of massive MIMO with MRC receivers against the number of BS antennas ($P=12$ and $K = 6$).
\vspace{-6mm}
\label{fig:MRC}}
\end{figure}

The simulation results and their corresponding analytical approximations of space-constrained massive MIMO systems with MRC are plotted in Fig. \ref{fig:MRC}. It is easily seen that the sum SE saturates with an increasing number of BS antennas for different total antenna array spaces $d_0$. This observation is consistent with \cite{masouros2015space} and showcases that MRC suffers a substantial performance degradation when spatial correlation is high (small $d_0$). Moreover, for the same number of BS antennas, a monotonic increase in the sum SE is achieved as $d_0$ becomes larger. We also observe that the gap between the curves decreases as $d_0$ increases, which implies that the effect of constrained space becomes less pronounced.

\section{ZF Receivers}\label{se:ZF}
We now turn our attention to the case of ZF receivers, which seek to eliminate inter-user interferences in massive MIMO systems. Let us consider the concept of ZF reception in \eqref{eq:received_signal} to obtain the ZF filter matrix ${\bf{T}}={\bf{G}}({\bf{G}}^H{\bf{G}})^{-1}$ in \eqref{eq:receivers}.

Then, the sum SE of ZF receivers can be expressed as
\begin{align}\label{eq:achievable_SE_zf_sum}
{R^{\text{ZF}}} = \sum\limits_{k = 1}^K {{\tt{E}}\left\{ {{{\log }_2}\left( {1 + \frac{{{p_u}}}{{{{\left[ {{{\left( {{{\bf{G}}^H}{\bf{G}}} \right)}^{ - 1}}} \right]}_{kk}}}}} \right)} \right\}} .
\end{align}
Next, we introduce a very tight lower bound on the achievable sum SE of ZF linear receivers \eqref{eq:achievable_SE_zf_sum}.

\subsection{Lower Bound}
\begin{prop}\label{prop:zf_lower}
For space-constrained massive MIMO systems with ZF receivers, the achievable sum SE is lower bounded as in \eqref{eq:achievable_SE_zf_lower_result} at the bottom of this page, where $\psi(\cdot)$ is the digamma function \cite[Eq. (8.36)]{gradshtein2000table}, and ${{{\bf{Y}}_n}}$ denotes a $P\times P$ matrix whose entries are
\setcounter{equation}{12}
\begin{equation}\label{eq:expected log_determinant_Y}
{\left[ {{{\bf{Y}}_n}} \right]_{p,q}} = \left\{
{\begin{array}{{lcr}}
{\beta _p^{q - 1},\;\;\;q\ne n},\\
{\beta _p^{q - 1}\ln{\beta _p},\;\;\;q = n}.
\end{array}} \right.
\end{equation}
\end{prop}
\begin{IEEEproof}
See Appendix \ref{se:zf_lower}.
\end{IEEEproof}


\subsection{Upper Bound}
We now move to the upper bound analysis, and present the following proposition.

\begin{prop}\label{prop:zf_upper}
For space-constrained massive MIMO systems with ZF receivers, the achievable sum SE is upper bounded as
\begin{align}\label{eq:achievable_SE_zf_upper_1_result}
{R_{\text{U}}^{\text{ZF}}}\leq{R_{\text{U}}^{\text{ZF}}}  &= K{\log _2}\Bigg( \frac{{ \left| {{{\bf{\Delta }}_2}} \right|}}{{\prod\nolimits_{i = 1}^{K - 1} {\Gamma \left( {K - i} \right)\prod\nolimits_{i < j}^{P} {\left( {{\beta _j} - {\beta _i}} \right)} } }} \notag \\
&+ {p_u}\frac{{ \left| {{{\bf{\Delta }}_1}} \right|}}{{\prod\nolimits_{i = 1}^K {\Gamma \left( {K - i + 1} \right)\prod\nolimits_{i < j}^P {\left( {{\beta _j} - {\beta _i}} \right)} } }} \Bigg) \notag \\
&- \frac{K}{{\ln 2}}\left( {\sum\limits_{n = 1}^{K - 1} {\psi \left( n \right) + \frac{{\sum\limits_{n = P - K + 2}^P {\left| {{{\bf{Y}}_n}} \right|} }}{{\prod\nolimits_{i < j}^P {\left( {{\beta _j} - {\beta _i}} \right)} }}} } \right),
\end{align}
where $\Gamma(\cdot)$ denotes the Gamma function \cite[Eq. (8.31)]{gradshtein2000table}, ${{\bf{\Delta }}_1} = [{{\bf{\Xi}}_1}{{\bf{ \Phi}}_1 }]$ is a $P \times P$ matrix with entries
\begin{align}
{\left[ {\bf{\Xi }}_1 \right]_{p,q}} &= \beta _p^{q - 1},\;\;\;q = 1,2, \dots ,P - K,\notag \\
{\left[ {\bf{\Phi }}_1 \right]_{p,q}} &= \beta _p^{q }\Gamma \left( q-P+K+1 \right),\;\;\;q= P-K+1, \dots ,P,\notag
\end{align}
and
${{\bf{\Delta }}_2} = [{{\bf{\Xi}}_2}{{\bf{ \Phi}}_2 }]$ is a $P \times P$ matrix with entries
\begin{align}
{\left[ {\bf{\Xi }}_2 \right]_{p,q}} &= \beta _p^{q - 1},\;\;\;q = 1,2, \dots ,P - K+1,\notag \\
{\left[ {\bf{\Phi }}_2 \right]_{p,q}} &=\beta _p^{q  }\Gamma \left( q-P+K \right),\;\;\;q= P-K+2, \dots ,P.\notag
\end{align}
\end{prop}
\begin{IEEEproof}
See Appendix \ref{se:zf_upper}.
\end{IEEEproof}

\begin{figure}[htbp]
\centering
\includegraphics[scale=0.6]{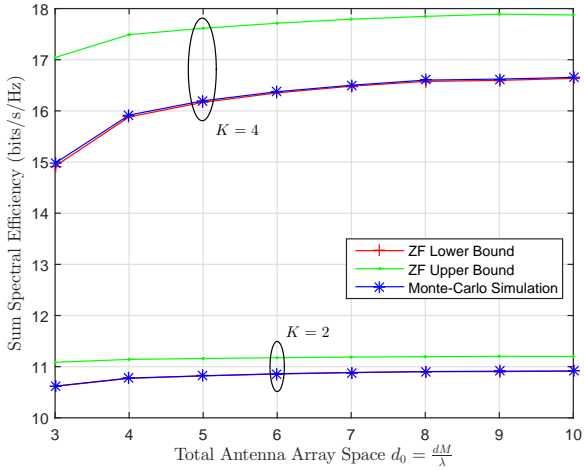}
\caption{Simulated and analytical approximation of the sum SE of massive MIMO with ZF receivers against the total antenna array space $d_0=\frac{dM}{\lambda}$ ($M=100$ and $P= 12$).
\vspace{-4mm}
\label{fig:zf_lower_upper_bound_d}}
\end{figure}

In Figs. \ref{fig:zf_lower_upper_bound_K} and \ref{fig:zf_lower_upper_bound_d}, the simulated achievable sum SE along with the proposed lower bound \eqref{eq:achievable_SE_zf_lower_result} and upper bound \eqref{eq:achievable_SE_zf_upper_1_result} are plotted against the number of BS antennas and total antenna array space, respectively. Clearly, all lower bounds can predict
the exact sum SE for all the considered cases, which validate their tightness. On the other hand, the upper bounds are relatively looser, due to the large variance of the involved random variables. Figure \ref{fig:zf_lower_upper_bound_K} indicates that adding more antennas significantly improves the sum SE of the massive MIMO link by suppressing thermal noise, even in the space constrained scenario. Moreover, from Fig. \ref{fig:zf_lower_upper_bound_d}, we observe that the SE does improve with increased total physical space, particularly for the case of more UEs.

\begin{figure}[t]
\centering
\includegraphics[scale=0.6]{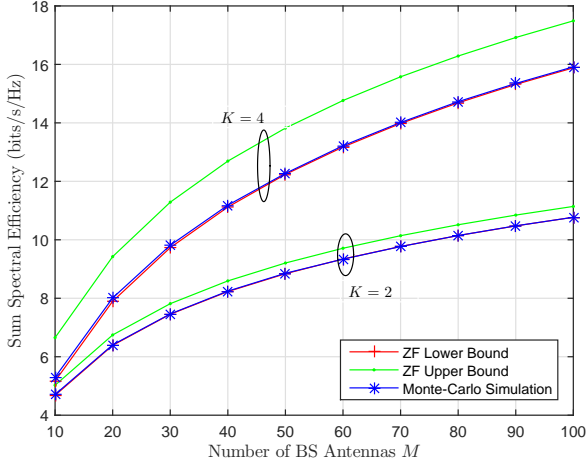}
\caption{Simulated and analytical approximation of the sum SE of massive MIMO with ZF receivers against the number of BS antennas ($P=12$ and $d_0 = 4$).
\vspace{-4mm}
\label{fig:zf_lower_upper_bound_K}}
\end{figure}

\section{MMSE Receivers}\label{se:MMSE}
For MMSE receivers, the receiver matrix ${\bf{T}}$ is given by \cite{ngo2013energy}
\begin{align}
{{\bf{T}}^H} = {\left( {{{\bf{G}}^H}{\bf{G}} + \frac{1}{{{p_u}}}{{\bf{I}}_K}} \right)^{ - 1}}{{\bf{G}}^H} = {{\bf{G}}^H}{\left( {{\bf{G}}{{\bf{G}}^H} + \frac{1}{{{p_u}}}{{\bf{I}}_M}} \right)^{ - 1}}.\notag
\end{align}
The achievable sum SE can be written as
\begin{align}
{R^{\text{MMSE}}} &= \sum\limits_{k = 1}^K {{\tt{E}}\left\{ {{{\log }_2}\left( {\frac{1}{{{{\left[ {{{\left( {{{\bf{I}}_K} + {p_u}{{\bf{G}}^H}{\bf{G}}} \right)}^{ - 1}}} \right]}_{kk}}}}} \right)} \right\}} \label{eq:mmse_sum_SE_1}\\
& = K{\tt{E}}\left\{ {{{\log }_2}\left( {\left| {{{\bf{I}}_K} + {p_u}{{\bf{G}}^H}{\bf{G}}} \right|} \right)} \right\}\notag \\
& - \sum\limits_{k = 1}^K {{\tt{E}}\left\{ {{{\log }_2}\left( {\left| {{{\bf{I}}_{K - 1}} + {p_u}{\bf{G}}_k^H{{\bf{G}}_k}} \right|} \right)} \right\}} \label{eq:mmse_sum_SE_2},
\end{align}
where \eqref{eq:mmse_sum_SE_2} can be derived from \eqref{eq:mmse_sum_SE_1} with the aid of an important matrix property \cite[Eq. (11)]{matthaiou2011novel} as
\begin{align}\label{eq:det_matrix_property}
{\left[ {{{\left( {{{\bf{G}}^H}{\bf{G}}} \right)}^{ - 1}}} \right]_{kk}} = \frac{{\left| {{\bf{G}}_k^H{{\bf{G}}_k}} \right|}}{{\left| {{{\bf{G}}^H}{\bf{G}}} \right|}}.
\end{align}

The following proposition presents an exact closed-form expression for the achievable sum SE of MMSE receivers.

\begin{prop}\label{prop:mmse_exact}
For space-constrained massive MIMO systems with MMSE receivers, the exact sum SE is given by
\begin{align}\label{eq:mmse_sum_SE_4}
{R^{\text{MMSE}}} &= \frac{{K{{\log }_2}e}}{{\prod\nolimits_{i < j}^P {\left( {{\beta _j} - {\beta _i}} \right)} }}\sum\limits_{l = 1}^P \sum\limits_{n = P - K + 1}^P \beta _l^{n - 1}{e^{1/{\beta _l}{p_u}}}\notag \\
&\times{D_{l,n}}{E_{n - P + K}}\left( {\frac{1}{{{\beta _l}{p_u}}}} \right) ,
\end{align}
where ${D_{l,n}}$ is the $(l,n)$th cofactor of a $P\times P$ matrix $\bf{D}$ with the $(p,q)$th entry $[{\bf{D}}]_{p,q}=\beta_p^{q-1}$, and $E_x(y)$ is the exponential integral function \cite{gradshtein2000table}.
\end{prop}
\begin{IEEEproof}
See Appendix \ref{se:mmse_exact}.
\end{IEEEproof}

\begin{figure}[t]
\centering
\includegraphics[scale=0.6]{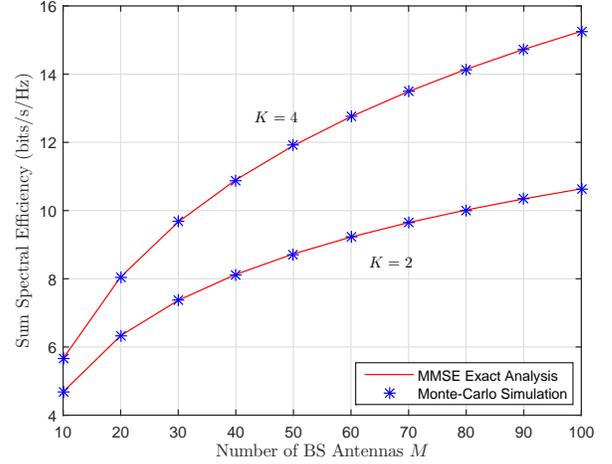}
\caption{Simulated and analytical expression of the sum SE of massive MIMO with MMSE receivers against the number of antennas at BS ($P=12$ and $d_0 = 4$).
\vspace{-4mm}
\label{fig:MMSE_K}}
\end{figure}
\begin{figure}[htbp]
\centering
\includegraphics[scale=0.6]{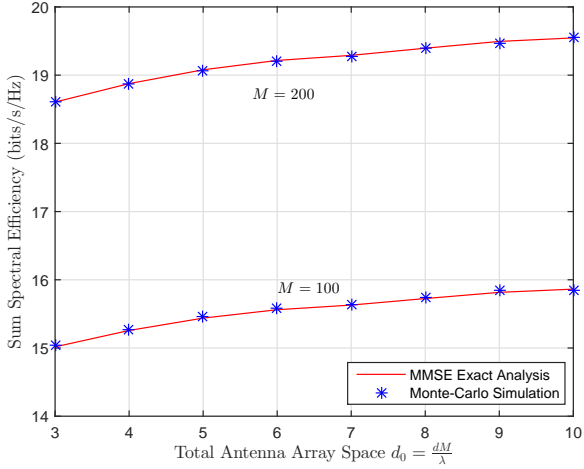}
\caption{Simulated and analytical expression of the sum SE of massive MIMO with MMSE receivers against the total antenna array space $d_0=\frac{dM}{\lambda}$ ($K=4$ and $P= 8$).
\vspace{-4mm}
\label{fig:MMSE_distance_M}}
\end{figure}

For MMSE receivers, Figs. \ref{fig:MMSE_K} and \ref{fig:MMSE_distance_M} investigate the simulated and analytical sum SE of space-constrained massive MIMO systems against the number of BS antennas and the total antenna array space. It is clear to see that the exact analytical results are indistinguishable from the numerical simulations, which validates the correctness of the derived expressions.
Furthermore, Fig. \ref{fig:MMSE_distance_M} reveals that with a fixed total antenna array space, the sum SE can be still increased by employing more BS antennas. This is because the improved array gain caused by the increased $M$ dominates the sum SE loss due to the reduced $d_0$.

\section{Conclusions}
In this paper, we investigated the performance of massive MIMO systems with a practical space-constrained topology, where the antenna array at the BS has a limited total space. This introduces an increasing spatial correlation with an increased number of BS antennas. We first derived the approximated sum SE with MRC receivers. Through analytical and numerical results, we confirmed that a saturation of the achievable sum SE occurs with an increasing number of BS antennas. For ZF receivers, we derived new lower and upper bounds on the sum SE, which increases for a higher number of UEs, as long as $M \gg K$. Moreover, the proposed lower bound is tighter than the upper bound. For MMSE receivers, an exact expression for the sum SE is derived and validated by simulation results, which shows that the sum SE increases with the number of BS antennas. This is due to the fact that the signal-to-interference-plus-noise ratios (SINRs) of ZF and MMSE receivers increase with the number of BS antennas, while MRC receivers can only work well at low SINRs.

\begin{appendices}

      \section{Proof of Proposition 1}\label{se:A}
By employing \textit{Lemma 1} in \cite{zhang2014power}, the approximated ${{ {R}}_k^{\text{MRC}}}$ can be expressed as
\setcounter{equation}{23}
\begin{align}\label{eq:achievable_SE_appro}
{{ {R}}_k^{\text{MRC}}} \approx {\log _2}\left( {1 + \frac{{{p_u}{{\tt{E}}}\left\{ {{{\left\| {{{\bf{g}}_k}} \right\|}^4}} \right\}}}{{{p_u}\sum\nolimits_{l \ne k}^K {{{\tt{E}}}\left\{ {{{\left| {{\bf{g}}_k^H{{\bf{g}}_l}} \right|}^2}} \right\}}  + {{\tt{E}}}\left\{ {{{\left\| {{{\bf{g}}_k}} \right\|}^2}} \right\}}}} \right).
\end{align}
From \textit{Lemma 2} of \cite{bjornson2014massive}, the numerator term of \eqref{eq:achievable_SE_appro} can be calculated as
\begin{align}\label{eq:numerator}
{\tt{E}}\left\{ {{{\left\| {{{\bf{g}}_k}} \right\|}^4}} \right\}&= {\zeta_k^2}{\left| {{\text{tr}}\left\{ {{{\bf{A}}^H}{\bf{A}}} \right\}} \right|^2} + {\zeta_k^2} {{\text{tr}}}\left\{ {{{\left( {{{\bf{A}}^H}{\bf{A}}} \right)}^2}} \right\}  \notag \\
&= {\zeta_k^2}\left(M^2 + \sum\limits_{i = 1}^P {\beta _i^2}\right).
\end{align}
Note that ${\bf{Z}} \in \mathbb{C}^{P\times P}$ is a deterministic matrix. Considering the definition of $\bf{A}$ and ${\bf{h}}_k$ and using \textit{Lemma 2} of \cite{bjornson2014massive} again, the first term in the denominator of \eqref{eq:achievable_SE_appro} can be derived as
\begin{align}\label{eq:denominator1}
{\tt{E}}\left\{ {{{\left| {{\bf{g}}_k^H{{\bf{g}}_l}} \right|}^2}} \right\} = {\zeta_k \zeta_l}{\tt{E}}\left\{{ \left| {{\bf{h}}_k^H{{\bf{A}}^H}{\bf{A}}{{\bf{h}}_l}}\right|}^2 \right\} = {\zeta_k \zeta_l}\sum\limits_{i = 1}^P {\beta _i^2} ,
\end{align}
and the second term in the denominator of \eqref{eq:achievable_SE_appro} is given by
\begin{align}\label{eq:denominator2}
{\tt{E}}\left\{ {{{\left\| {{{\bf{g}}_k}} \right\|}^2}} \right\} &= {\zeta_k^2} {\tt{E}}\left\{ {{\bf{h}}_k^H{{\bf{A}}^H}{\bf{A}}{{\bf{h}}_k}} \right\}= \sum\limits_{i = 1}^P {{\zeta_k^2} \beta _i } = M{\zeta_k^2}.
\end{align}
Substituting \eqref{eq:numerator}, \eqref{eq:denominator1}, and \eqref{eq:denominator2} into \eqref{eq:achievable_SE_appro}, we can derive \eqref{eq:achievable_SE_appro_result} in Proposition 1.
      \section{Proof of Proposition 2}\label{se:zf_lower}
We start from \eqref{eq:achievable_SE_zf_sum}, and apply Jensen's inequality on the convex function $\log_2\left({1+a\exp(x)}\right)$ for $a>0$ to get
\begin{align}
&{R^{\text{ZF}}}\geqslant {R_{\text{L}}^{\text{ZF}}}= \notag \\
&\sum\limits_{k = 1}^K {{{\log }_2}\left( {1 + {p_u}e^{ \left( {{\tt{E}}\left\{ {\ln \left( {\frac{1}{{{{\left[ {{{\left( {{{\bf{G}}^H}{\bf{G}}} \right)}^{ - 1}}} \right]}_{kk}}}}} \right)} \right\}} \right)}} \right)}    \label{eq:achievable_SE_zf_lower_1}\\
&= \sum\limits_{k = 1}^K {{{\log }_2}\left( {1 + {p_u} e^{ \left( {{\tt{E}}\left\{ {\ln \left( {\left| {{{\bf{G}}^H}{\bf{G}}} \right|} \right)} \right\} - {\tt{E}}\left\{ {\ln \left( {\left| {{\bf{G}}_k^H{{\bf{G}}_k}} \right|} \right)} \right\}} \right)}} \right)} ,\label{eq:achievable_SE_zf_lower_2}
\end{align}
where from \eqref{eq:achievable_SE_zf_lower_1} to \eqref{eq:achievable_SE_zf_lower_2}, we have used \eqref{eq:det_matrix_property}.
By utilizing \textit{Lemma 4} of \cite{jin2010ergodic}, the average log-determinant of ${{{\bf{G}}^H}{\bf{G}}}$ can be derived as
\begin{align}\label{eq:expected log_determinant}
{\tt{E}}\left\{ {\ln \left( {\left| {{{\bf{G}}^H}{\bf{G}}} \right|} \right)} \right\} &= \left(\sum\limits_{n = 1}^K {\psi \left( n \right)} + \frac{{\sum\limits_{n = P - K + 1}^P {\left| {{{\bf{Y}}_n}} \right|} }}{{\prod\nolimits_{i < j}^P {\left( {{\beta _j} - {\beta _i}} \right)} }} \right) \notag \\
&\times \ln {\sum\limits_{n = 1}^K {\zeta_n}}.
\end{align}
Note that ${{\bf{G}}_k}$ is an $M\times (K-1)$ matrix, and we have
\begin{align}\label{eq:expected log_determinant_k}
{\tt{E}}\left\{ {\ln \left( {\left| {{{{\bf{G}}_k^H}}{{\bf{G}}_k}} \right|} \right)} \right\} &= \left(\sum\limits_{n = 1}^{K-1} {\psi \left( n \right) + \frac{{\sum\limits_{n = P - K + 2}^P {\left| {{{\bf{Y}}_n}} \right|} }}{{\prod\nolimits_{i < j}^P {\left( {{\beta _j} - {\beta _i}} \right)} }}}   \right)\notag \\
&\times \ln {\sum\limits_{n=1,n \neq k}^K {\zeta_n}}.
\end{align}
Substituting \eqref{eq:expected log_determinant} and \eqref{eq:expected log_determinant_k} into \eqref{eq:achievable_SE_zf_lower_2}, we can complete the proof of Proposition 2.
\section{Proof of Proposition 3}\label{se:zf_upper}
By applying \eqref{eq:det_matrix_property} and Jensen's inequality again, we derive the upper bound $R_{U}$ on the uplink sum SE \eqref{eq:achievable_SE_zf_sum} as
\begin{align}\label{eq:achievable_SE_zf_upper_2}
{R^{\text{ZF}}}\leqslant{R_{\text{U}}^{\text{ZF}}} &= \sum\limits_{k = 1}^K {{{\log }_2}\left( {{{\tt{E}}}\left\{ { \left| {{\bf{G}}_k^H{{\bf{G}}_k}} \right|} \right\} + {p_u}{{\tt{E}}}\left\{ { \left| {{{\bf{G}}^H}{\bf{G}}} \right|} \right\}} \right)}  \notag \\
&- \sum\limits_{k = 1}^K {{{\tt{E}}}\left\{ {{{\log }_2}\left( { \left| {{\bf{G}}_k^H{{\bf{G}}_k}} \right|} \right)} \right\}}.
\end{align}
In order to obtain ${R_{\text{U}}^{\text{ZF}}}$, we first need to derive ${\tt{E}}\left\{ { \left| {{{\bf{G}}^H}{\bf{G}}} \right|} \right\}$. Note that the joint probability density function (PDF) of the unordered eigenvalues $\tau_1, \tau_2, \cdots, \tau_K$ of ${{{\bf{G}}^H}{\bf{G}}}$ is given by \cite[Eq. (86)]{jin2010ergodic}
\begin{align}\label{eq:joint_pdf_eigenvalues}
f\left( {{\tau _1}, \cdots ,{\tau _K}} \right) = \frac{{\left| {\bf{\Delta }} \right|\prod\nolimits_{i < j}^K {\left( {{\tau _j} - {\tau _i}} \right)} }}{{K\prod\nolimits_{i = 1}^K {\Gamma \left( {K - i + 1} \right)\prod\nolimits_{i < j}^P {\left( {{\beta _j} - {\beta _i}} \right)} } }} ,
\end{align}
where ${\bf{\Delta }}$ is the $P\times P$ matrix given by
\begin{align}
{\bf{\Delta }} = \left[ {\begin{array}{*{20}{c}}
1&  \cdots  &{\beta _1^{P\! -\! K\! -\! 1}{e^{ - {\tau _1}/{\beta _1}}}}& \cdots &{\beta _1^{P\! -\! K \!- \!1}{e^{ - {\tau _K}/{\beta _1}}}}\\
 \vdots  & \ddots   & \vdots & \ddots & \vdots \\
1 & \cdots  &{\beta _P^{P \!-\! K\! -\! 1}{e^{ - {\tau _1}/{\beta _P}}}}& \cdots &{\beta _P^{P \!-\! K \!-\! 1}{e^{ - {\tau _K}/{\beta _P}}}}
\end{array}} \right].
\end{align}
Substituting \eqref{eq:joint_pdf_eigenvalues} into ${{\tt{E}}}\left\{ { \left| {{{\bf{G}}^H}{\bf{G}}} \right|} \right\}$, we can obtain
\begin{align}\label{eq:expected_determinant}
&{{\tt{E}}}\left\{ {\left| {{{\bf{G}}^H}{\bf{G}}} \right|} \right\}= {\tt{E}}\left\{ {\prod\limits_{i = 1}^K {{\tau _i}} } \right\} \notag \\
&= \int_{\tt{D_{ord}}}  {   {\frac{{\left| {\bf{\Delta }} \right|\prod\limits_{i = 1}^K {{\tau _i}} \prod\nolimits_{i < j}^K {\left( {{\tau _j} - {\tau _i}} \right)} }}{{K\prod\nolimits_{i = 1}^K {\Gamma \left( {K - i + 1} \right)\prod\nolimits_{i < j}^P {\left( {{\beta _j} - {\beta _i}} \right)} } }}} {d{{\tau_1}}} \dots {d{{\tau_K}}}},
\end{align}
where ${\tt{D_{ord}}}=\left\{ \infty \geq \tau_1 \geq \dots \geq \tau_K \right\}$ is the integration region. Applying the integral identity from \cite[Lemma 2]{shin2006capacity} and \cite[Eq. (3.351.3)]{gradshtein2000table}, \eqref{eq:expected_determinant} can be evaluated in closed-form as
\begin{align}\label{eq:expected_determinant_result}
{{\tt{E}}}\left\{ {\left| {{{\bf{G}}^H}{\bf{G}}} \right|} \right\} = \frac{{ \left| {{{\bf{\Delta }}_1}} \right|}}{{\prod\nolimits_{i = 1}^K {\Gamma \left( {K - i + 1} \right)\prod\nolimits_{i < j}^P {\left( {{\beta _j} - {\beta _i}} \right)} } }},
\end{align}
while
\begin{align}\label{eq:expected_determinant_result_2}
{{\tt{E}}}\left\{ {\left| {{\bf{G}}_k^H{{\bf{G}}_k}} \right|} \right\} = \frac{{ \left| {{{\bf{\Delta }}_2}} \right|}}{{\prod\nolimits_{i = 1}^{K-1} {\Gamma \left( {K - i} \right)\prod\nolimits_{i < j}^{P} {\left( {{\beta _j} - {\beta _i}} \right)} } }}.
\end{align}
Combining \eqref{eq:expected log_determinant_k}, \eqref{eq:expected_determinant_result}, and  \eqref{eq:expected_determinant_result_2}, we can derive the upper bound in Proposition \ref{prop:zf_upper}.
\section{Proof of Proposition 4}\label{se:mmse_exact}
Considering the unified PDF expression of the unordered eigenvalue $\tau$ of an $M\times K$ complex semi-correlated central Wishart matrix with $K$ degrees of freedom from \cite[Eq. (14)]{jin2010ergodic}
\begin{align}\label{eq:PDF_eigenvalue}
{f_\tau }\left( x \right) &= \frac{1}{{K\prod\nolimits_{i < j}^P {\left( {{\beta _j} - {\beta _i}} \right)} }}\notag\\
&\times\sum\limits_{l = 1}^P {\sum\limits_{n = P-K+1}^P {\frac{{{x^{K + n - P - 1}}{e^{ - x/{\beta _l}}}\beta _l^{P - K - 1}}}{{\Gamma \left( { K-P + n} \right)}}} } {D_{l,n}}.
\end{align}
Substituting \eqref{eq:PDF_eigenvalue} into \eqref{eq:mmse_sum_SE_2}, and using the integral identity \cite{alfano2004mutual}, we can derive the exact sum SE with MMSE receivers as
\begin{align}\label{eq:mmse_sum_SE_3}
{R^{\text{MMSE}}}&= \frac{{K{{\log }_2}e}}{{\prod\nolimits_{i < j}^P {\left( {{\beta _j} - {\beta _i}} \right)} }}\sum\limits_{l = 1}^P {e^\frac{1}{{\beta _l}{p_u}}} \notag \\
&\times \Bigg( \sum\limits_{n = P - K + 1}^P \beta _l^{n - 1}{D_{l,n}}\sum\limits_{h = 1}^{K + n - P}{E_h}\left( {\frac{1}{{{\beta _l}{p_u}}}} \right)\notag \\
& - \sum\limits_{n = P - K + 2}^P {\beta _l^{n - 1}{D_{l,n}}\sum\limits_{h = 1}^{K + n - P - 1} {{E_h}\left( {\frac{1}{{{\beta _l}{p_u}}}} \right)} }    \Bigg).
\end{align}
After some tedious but straightforward manipulations, the proof can be completed.
\end{appendices}

\bibliographystyle{IEEEtran}
\bibliography{IEEEabrv,Ref}

\end{document}